\documentclass[nonacm, sigconf, screen]{acmart}

\usepackage{amsmath,amsfonts}

\usepackage{amssymb}
\usepackage{algorithmic}
\usepackage{graphicx}
\usepackage{textcomp}
\usepackage{xcolor}
\usepackage{url}
\usepackage{booktabs}
\usepackage{makecell}
\usepackage{multirow}
\usepackage{multicol}

\AtBeginDocument{%
  }

\begin{document}

\title[Overcoming Quadratic Hardware Scaling for a Fully Connected Digital Oscillatory Neural Network]{Overcoming Quadratic Hardware Scaling for a\\Fully Connected Digital Oscillatory Neural Network}

\author{Bram Haverkort}
\email{b.f.haverkort@tue.nl}
\author{Aida Todri-Sanial}
\email{a.todri.sanial@tue.nl}
\affiliation{%
  \institution{Eindhoven University of Technology}
  \city{Eindhoven}
  \country{The Netherlands}
}


\begin{abstract}
  Computing with coupled oscillators or oscillatory neural networks (ONNs) has recently attracted a lot of interest due to their potential for massive parallelism and energy-efficient computing.
However, to date, ONNs have primarily been explored either analytically or through analog circuit implementations.
This paper shifts the focus to the digital implementation of ONNs, examining various design architectures.
We first report on an existing digital ONN design based on a recurrent architecture.
The major challenge for scaling such recurrent architectures is the quadratic increase in coupling hardware with the network size.
To overcome this challenge, we introduce a novel hybrid architecture that balances serialization and parallelism in the coupling elements that shows near-linear hardware scaling, on the order of about 1.2 with the network size.
Furthermore, we evaluate the benefits and costs of these different digital ONN architectures in terms time to solution and resource usage on FPGA emulation.
The proposed hybrid architecture allows for a 10.5$\times$ increase in the number of oscillators while using 5-bits to represent the coupling weights and 4-bits to represent the oscillator phase on a Zynq-7020 FPGA board.
The near-linear scaling is a major step towards implementing large scale ONN architectures.
To the best of our knowledge, this work presents the largest fully connected digital ONN architecture implemented thus far with a total of 506 fully connected oscillators.
\end{abstract}


\maketitle

\section{Introduction} \label{sec:introduction}
With the growing computational demands and the rise of AI, we are witnessing a significant increase in the power consumption of computing systems, and this growth is becoming unsustainable, particularly with the widespread deployment of smart systems.
One of the main reasons for this trend is the computing architecture based on the von Neumann paradigm, which has inherent limitations due to the separation of memory and processing units, leading to memory and power wall bottlenecks.
To overcome these limitations, significant efforts in recent years are focused on neuromorphic computing architectures, such as spiking neural networks (SNNs) \cite{maass_networks_1997, yao_spike-based_2024}.
Inspired by biological neural networks for their energy efficiency, these efforts aim to develop architectures that eliminate the separation between memory and processing, moving towards an in-memory computing paradigm and thereby reducing the need for data shuffling between memory and processing units.

\begin{figure}[!t]
    \centering
    \includegraphics[width=\linewidth]{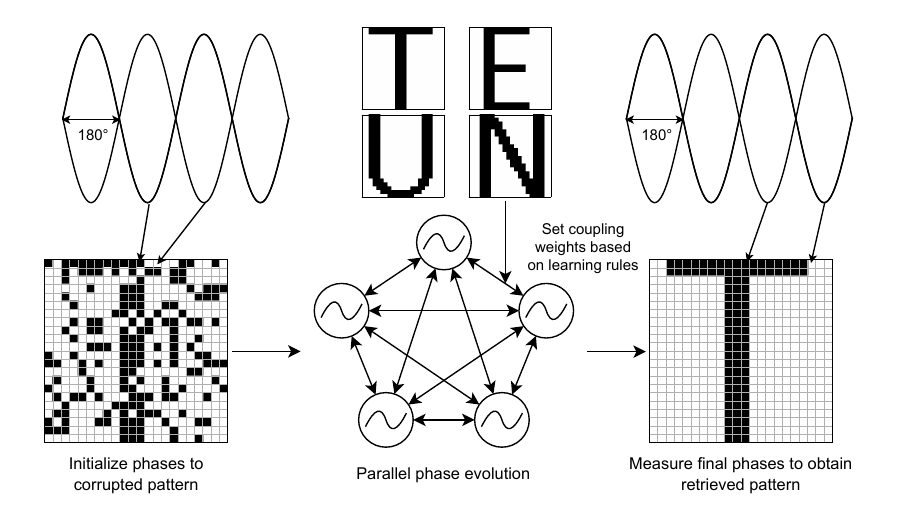}
    \caption{Illustration of oscillatory neural network for pattern retrieval. Pattern retrieval process where each pixel in the pattern is represented with an oscillator. The coupling weights in the network are determined using a learning rule. A corrupted pattern is injected into a network as initial condition. The oscillator phases naturally evolve in parallel to the memorized pattern.}
    \label{fig:bg_generic_onn}
\end{figure}

In this work, we present a novel physics-inspired computing paradigm based on coupled oscillators.
This approach not only eliminates the separation between memory and processing but also enables energy-efficient computing by harnessing complex dynamics to achieve massive parallelism, distinguishing it from other neuromorphic computing paradigms \cite{todri-sanial_computing_2024}.
Coupled oscillator networks are typically structured as recurrent networks that perform computations in phase and frequency domain contrary to other neural networks that compute in time domain.
Coupled oscillators or Oscillatory Neural Networks (ONNs) originate from Hopfield neural networks (HNNs) \cite{hopfield_neural_1982, ramsauer_hopfield_2021}, illustrated in Figure \ref{fig:bg_generic_onn}.
Unlike Hopfield Neural Networks (HNNs) that use binary neurons (e.g., spin-up +1, spin-down -1), ONNs can surpass binary limitations. In ONNs, information is encoded in phase, providing additional degrees of freedom to represent data within a phase space.
For instance, a spin-up state can be represented by a 0° phase, while a spin-down state corresponds to a 180° phase.
Additionally, finer phase differences can encode even greater amounts of information.
Like HNNs, ONNs are energy-minimizing networks that inherently perform associative memory tasks, such as pattern recognition
\cite{hoppensteadt_pattern_2000, abernot_digital_2021, abernot_oscillatory_2023, delacour_oscillatory_2021}.
ONNs have also been adapted and shown promising performance for image classification \cite{abernot_training_2023, sabo_classonn_2024}.
Typically, these applications are tackled using convolutional neural networks \cite{stanojevic_high-performance_2024, syed_exploring_2021} and spiking neural networks \cite{byerly_no_2021, wu_improvement_2021}.
However, these networks often feature high complexity, using many layers and parameters, with numbers in the tens and millions, respectively \cite{simonyan_very_2015}.
Additionally, the energy minimization of coupled networks of oscillators can be described by the Ising model.
This model applied to ONNs has garnered significant interest in recent years for designing oscillatory Ising machines, which can be utilized to solve various complex optimization problems, such as max-cut, graph coloring, and more \cite{vaidya_creating_2022, delacour_mixed-signal_2023, wang_oim_2019, wang_solving_2021, bashar_designing_2023, bashar_fpga-based_2024, sundara_raman_sachi_2024, cilasun_3sat_2024, nikhar_all--all_2024, bashar_oscillator-based_2021}.
Since the number of oscillators in a network required to solve a certain problem is determined by the dataset or exact problem that needs to be embedded, this motivates the need for larger ONNs.
For example, solving the max-cut problem on a graph requires each graph node to be represented by one oscillator, or in the case of associative memory tasks, each oscillator is typically used to represent one pixel in a pattern.
Additionally, the network topology is also related to the ease of problem embedding.
Sparse network topologies such as nearest-neighbor or King's graphs do not support the embedding of all problems and can require an additional computational step to map a problem.
However, architectures using sparse networks typically have a higher number of nodes compared to architectures with fully-connected networks since sparse networks require fewer connections compared to a fully-connected topology.

Typically, ONNs are designed in the analog domain to emulate the complex dynamics of coupled oscillators.
Recently, the first digital ONN implementation was reported \cite{abernot_digital_2021,abernot_simulation_2023, abernot_training_2023}, featuring a recurrent ONN architecture that represents the dynamics of digital square-wave oscillators.
This recurrent architecture has been implemented in Field Programmable Gate Arrays (FPGA) and demonstrated ONNs for performing tasks such as pattern retrieval and edge detection.
While ONNs have been successfully demonstrated  on FPGA for their potential in edge computing, a significant limitation is the quadratic increase in the number of coupling elements as the network size grows.
Fully connected small ONNs are feasible, but scaling up to larger networks for solving realistic problems remains a challenge.
In this work, for the first time, we propose a novel ONN computing architecture that is not recurrent but a hybrid architecture that enables near-linear network scalability instead of quadratic scalability.
This hybrid architecture can be used to accelerate computing on the edge, or scaled up to solve large combinatorial optimization problems.

The rest of this paper is structured as follows.
Section \ref{sec:background} provides background on the computing principles of ONNs, state-of-the-art implementations and architectures of ONNs, and describes the existing recurrent ONN architecture.
Section \ref{sec:hybridarchitecture} details the proposed hybrid ONN architecture and its implementation details.
Section \ref{sec:methods} explains the methodology behind the architecture comparison, followed by 
the comparison between recurrent and hybrid ONN architectures in Section \ref{sec:results}.
The discussions are presented Section \ref{sec:discussion}.
Finally, Section \ref{sec:conclusion} concludes the paper.

\section{Background} \label{sec:background}
\subsection{Oscillatory Neural Networks} \label{sec:background_onn}
Oscillatory Neural Networks (ONNs) are fully connected networks of coupled oscillators based on Hopfield neural networks \cite{hopfield_neural_1982, ramsauer_hopfield_2021} and the Ising model \cite{ising_beitrag_1925}.
The network minimizes the Hamiltonian given by
\begin{equation}
    H = -\sum_{i,j}J_{ij}\sigma_i\sigma_j-\mu\sum_ih_i\sigma_i,
\end{equation}
where $J_{ij}$ is the coupling value between spins $\sigma_i$ and $\sigma_j$, which are represented by the oscillators, $\mu$ is the magnetic moment, and $h_i$ is the external magnetic field acting on spin $\sigma_i$ . The system naturally evolves to a configuration of spins $\sigma_i$ that minimizes $H$.
The dynamics in an oscillatory neural network can also be described by the dynamics in a network of phase-locked loop coupled oscillators given by \cite{hoppensteadt_pattern_2000}
\begin{equation}
    \dot{\theta}_i = \omega + V(\theta_i)\sum^{n}_{j=1}W_{ij}V\left(\theta_j - \frac{\pi}{2}\right),
\end{equation}
where $\omega$ is the natural frequency of the oscillator, $V(\theta)$ is the waveform function, and $W_{ij}$ is the coupling strength from oscillator $j$ to $i$.
In the case of a digital oscillator $V(\theta)$ represents a square signal based oscillator.
Similar to the Ising model, the oscillator phases naturally evolve to a state where the energy in the network is minimized.

A common application for ONNs is pattern retrieval using associative memory, where each oscillator represents one pixel in the pattern.
Here a dataset of patterns is embedded in the coupling weights of the network using a learning rule, for example the Diederich-Opper I learning rule \cite{diederich_learning_1987}.
After training the network, a corrupted pattern can be set as the initial condition for the phases of each oscillator.
The network will then naturally evolve to the closest learned pattern to reach a ground minima.
By measuring the final steady-state phases of the oscillators in relation to each other the retrieved pattern can be determined.
Figure \ref{fig:bg_generic_onn} shows this process.
By leveraging ONNs as oscillatory Ising machines, many other applications and algorithms are possible, including but not limited to: max-cut or max-k-cut \cite{bashar_fpga-based_2024}, traveling salesperson and image segmentation \cite{sundara_raman_sachi_2024}, and 3SAT or MaxSAT, also known as boolean satisfiability \cite{cilasun_3sat_2024, nikhar_all--all_2024, bashar_oscillator-based_2021}.

The approximate scaling in terms of number of network elements for a given number of oscillators is given in Table \ref{tab:bg_approximate_resources}.
\begin{table}
    \caption{Order of number of network elements for $N$ oscillators.}
    \label{tab:bg_approximate_resources}
    \begin{center}
        \begin{tabular}{|c|c|}
            \hline
            \textbf{Element} & \textbf{Order of scaling} \\ \hline \hline
            Oscillators & $N$ \\ \hline
            \makecell{Coupling elements} & $N^2$ \\ \hline
            \makecell{Memory cells for weights} & $N^2$ \\ \hline
        \end{tabular}
    \end{center}
\end{table}
In a fully connected architecture, each oscillator is connected to each other oscillator and itself, including self-coupling.
Hence, for $N$ oscillators each oscillator has $N$ connections and there will be $N^2$ connections in total.
Each connection is represented in hardware by one coupling element and one weight memory value, thus the number of coupling elements and memory cells are $N^2$.
The number of memory cells cannot be reduced for a given number of oscillators, because each weight value has to be stored in a memory cell.
It can be noted that one can cut the number of memory cells in half by assuming symmetric coupling in the network, but the order of the scaling will remain $N^2$, $\frac{N^2}{2}$ to be precise.
The architectures in this work allow for asymmetric coupling, thus the number of memory cells will be $N^2$.
The hardware resource usage of the individual oscillators can also be optimized, but since the coupling elements scale quadratically there is a higher return on investment to optimize the coupling elements compared to optimizing the oscillators themselves.
Since the number of memory cells cannot be reduced, that prompts one target for optimization, namely the coupling elements.
The coupling elements can be optimized in two ways.
One way is to optimize each individual coupling element to reduce the hardware resource usage per element, however this will not reduce the total scaling below $N^2$.
The second way is to share hardware resources between multiple coupling elements.
Hardware sharing potentially allows for drastic reduction in the number of total hardware elements and to go below quadratic scaling.
In this work the latter way of optimization of the coupling elements is explored.

\subsection{State-Of-The-Art Oscillatory Neural Networks and Oscillatory Ising Machines} \label{sec:background_sota}
We include an overview of the state of the art in Table \ref{tab:SOTA} of several digital ONNs and oscillatory Ising Machines, along with a few analog oscillator networks.
As can be seen from Table \ref{tab:SOTA}, there is a large variety of architectures with different number of nodes, number of connections, and network topologies.
A trade-off between the number of nodes and the network topology can be observed.
Architectures using sparse networks \cite{nikhar_all--all_2024, liu_1024-spin_2024, moy_1968-node_2022, wang_oim_2019, wang_solving_2021} typically have a higher number of nodes compared to architectures with all-to-all networks \cite{abernot_digital_2021, abernot_oscillatory_2023, abernot_training_2023, jackson_oscillatory_2018, luhulima_digital_2023, vaidya_creating_2022} since sparse networks require fewer connections compared to a fully-connected topology.
As can be seen, although the hybrid architecture introduced in this work, which is explained in later sections, is not competitive in the number of nodes, the number of connections is competitive and exceeded only by a digital stochastic differential equation (SDE) solver \cite{bashar_fpga-based_2024}, which operates on a different paradigm.
Note that the network topology is relevant for ease of problem embedding.
Sparse network topologies such as nearest-neighbor or King's graphs do not support the embedding of all problems and can require an additional computational step to map a problem.
An example is the embedding of a graph max-cut to a sparse network topology.
If one or more nodes of the graph has a degree greater than the number connections supported by the network topology, it cannot be embedded without applying a strategy like node merging, where multiple hardware nodes represent one graph node.
Hence, we choose to optimize an existing all-to-all recurrent ONN architecture, which allows for easy problem embedding.

\begin{table*}
    \caption{Comparison of oscillator-based architectures}
    \label{tab:SOTA}
    \centering
    \begin{tabular}{cccccc}
        \toprule
        \textbf{Reference} & \textbf{Oscillator} & \textbf{Nodes} & \textbf{Connection Type} & \textbf{Connections} & \textbf{Topology}\\
        \midrule
        \cite{abernot_digital_2021, abernot_oscillatory_2023, abernot_training_2023, luhulima_digital_2023} & Digital & 35 & Digital & 1190 & All-to-all \\
        \cite{jackson_oscillatory_2018} & Digital* & 100 & Analog (resistive) & 10000 & All-to-all \\
        \cite{nikhar_all--all_2024} & Digital P-bit & $1008^\text{¶}$ & Digital & $\approx9072^\text{§}$ & Neighbor + Configurable\\
        \cite{bashar_fpga-based_2024} & Digital SDE & $10000^\triangle$ & Digital & 80$^\dagger$ & All-to-all streamed \\
        \cite{liu_1024-spin_2024} & Ring Oscillator & 1024 & Analog (capacitive) & $\approx3716^\text{§}$ & King's graph \\
        \cite{moy_1968-node_2022} & Ring Oscillator & 1968 & Transmission Gates & $\approx7342^\text{§}$ & King's graph \\
        \cite{wang_oim_2019, wang_solving_2021} & Analog (LC) & 240 & Analog (resistive) & 1200 & 12 $\times$ 20 Chimera \\
        \cite{vaidya_creating_2022} & Analog (Schmitt Trigger) & 4 & 
 Analog (capacitive) & 6 & All-to-all\\
        This work (recurrent architecture) & Digital & 48 & Digital & 2256 & All-to-all\\
        This work (hybrid architecture) & Digital & 506 & Digital & 256036 & All-to-all serialized\\
        \bottomrule
    \end{tabular} \\
    \raggedright
    *: Authors indicate ``primarily digital'' \\
    ¶: Multiple configurations are shown. This is the highest number of spins of all configurations. \\
    §: Not reported: estimated based on network topology.\\
    $\triangle$: Virtual nodes, since it is a stochastic differential equation (SDE) accelerator.\\
    $\dagger$: States are streamed using chunks of size 80. One can argue that there are on the order of $10^8$ virtual connections.
    \label{tab:my_label}
\end{table*}

\subsection{Recurrent ONN Architecture} \label{sec:background_recurrent}
The architecture optimizations that will be discussed in this work are built on the recurrent digital ONN designed and used in \cite{abernot_digital_2021, abernot_oscillatory_2023, abernot_training_2023, luhulima_digital_2023}.
To establish a baseline, this architecture will be briefly reintroduced and explained.

\begin{figure}
  \includegraphics[width=\linewidth]{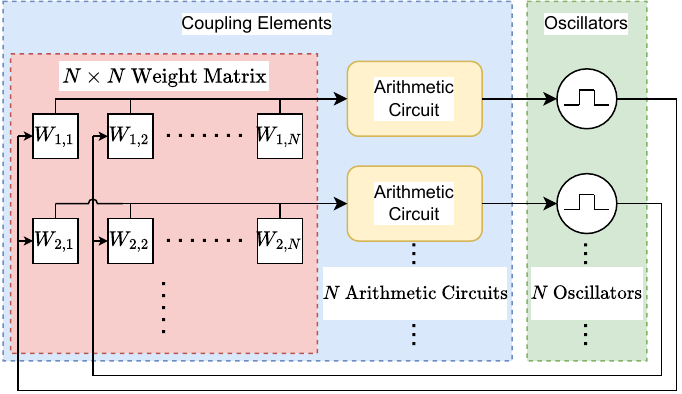}
  \caption{Global architecture overview for the digital ONN. The network is divided into two parts. One part for the coupling elements and one part for the oscillators. The coupling elements contain a $N\times N$ weight matrix for the coupling weights and $N$ arithmetic circuits, where $N$ is the number of oscillators in the architecture.}
  \label{fig:global_architecture}
\end{figure}

Figure \ref{fig:global_architecture} shows the global architecture overview for the digital ONN.
The network is divided into two main parts: On the left (in blue), the coupling elements and on the right (in green), the individual oscillators.

The coupling elements contain an $N\times N$ weight matrix, shown on the left side (in red), where $N$ is the number of oscillators, as well as an arithmetic circuit for each oscillator, shown on the right side (in yellow) in the coupling elements block, that computes the weighted sum of output signals of the other oscillators as in Figure \ref{fig:global_architecture}. 
The sign of the weighted sum is used to generate a reference signal, positive gives a high signal in the reference signal, negative gives a low signal in the reference signal.
If the weighted sum is exactly zero, the reference signal will match the current amplitude of the respective oscillator.
An edge detector and a counter measure the phase difference between the reference signal and the signal generated by the oscillator.
This phase difference is added to the phase of the respective oscillator to align the oscillator in phase with the reference signal.

\begin{figure}
  \includegraphics[width=\linewidth]{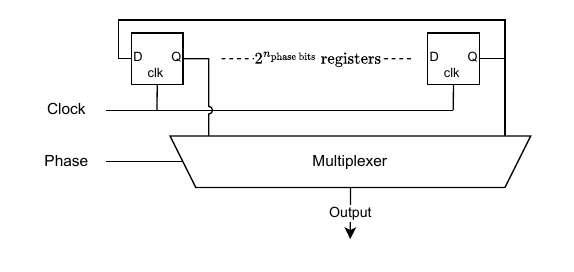}
  \caption{Phase controlled oscillator architecture. A circular shift registered is multiplexed to create a phase controlled oscillator. The first half of the registers are initialized with 1's and the second half with 0's.}
  \label{fig:oscillator_architecture}
\end{figure}

Each oscillator is implemented using a circular shift register-based phase-controlled oscillator, as shown in Figure \ref{fig:oscillator_architecture}.
By initializing the first half of the registers with value 1 and the second half with value 0 a square-wave oscillator is created.
The clock that drives the circular shift register can be considered to represent the natural frequency of the oscillator. 
After $2^{n_\text{phase bits}}$ clock cycles the oscillator will have completed one period, where $n_\text{phase bits}$ is the number of bits used to represent the phase of the oscillator.
Hence, the period of the oscillator is given by
\begin{equation}
    T_\text{oscillator} = 2^{n_\text{phase bits}}T_\text{clock}.
\end{equation}
The number of phase bits directly defines the number of circular shift register positions by 
\begin{equation}
    n_\text{registers} = 2^{n_\text{phase bits}}.
\end{equation}
Additionally, the number of phase bits also defines the size of the phase step by
\begin{equation}
    \text{size}_\text{phase step} = \frac{360\text{°}}{n_\text{registers}} = \frac{360\text{°}}{2^{n_\text{phase bits}}}.
\end{equation}
For example, if $n_\text{phase bits} = 4$, there will be 16 registers in the loop and the period of the oscillator will be 16 times larger than the clock period and the step size will be $360$°$/16 = 22.5$°.
For clarity, Table \ref{tab:shift_register} shows an example of the evolution of the register states over time for an oscillator with $n_\text{phase bits} = 2$.
Each row is an instance in time, measured in clock cycles, while each column is the value in each register.
At each time step the values in the shift registers are shifted to the left.
After 4 time steps the initial condition is reached, and one oscillation period has passed.
It can be seen that each column is a phase shifted version of the first column by one additional clock cycle.
Hence, by selecting a specific register using a multiplexer, a phase shift can be created for the oscillator.
\begin{figure*}
    \begin{minipage}[c]{0.475\linewidth}
        \includegraphics[width=\linewidth]{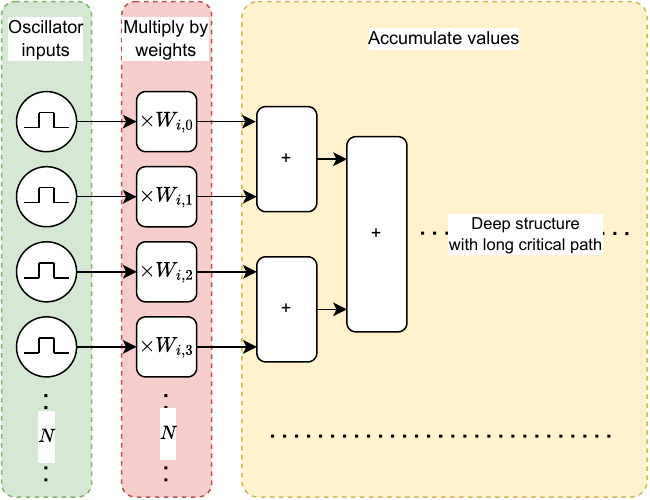}
        \caption{Parallel arithmetic circuit implementation for recurrent architecture. Oscillator amplitudes are multiplied by their respective coupling weight. These values are then accumulated to a total sum. The computation circuitry is implemented completely in combinatorial logic, leading to a deep logic structure with a long path delay.}
        \label{fig:adder_parallel}
    \end{minipage}
    \hfill
    \begin{minipage}[c]{0.475\linewidth}
        \includegraphics[width=\linewidth]{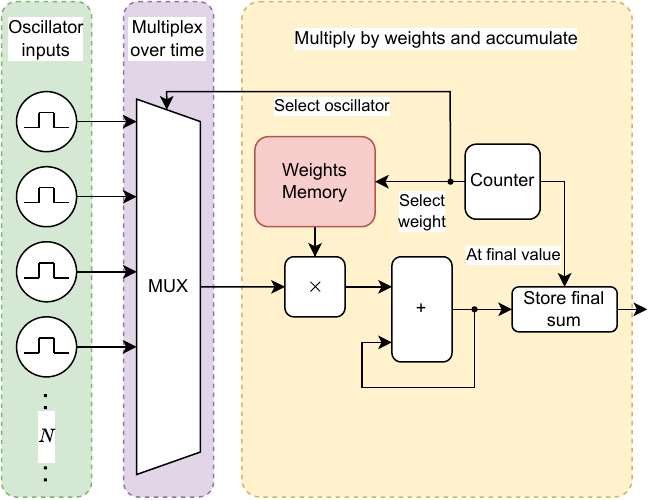}
        \caption{Serial arithmetic circuit implementation for hybrid architecture. Oscillator amplitudes are multiplied by their respective coupling weight. These values are then accumulated to a total sum. The computation circuitry is implemented with one adder to compute the sum serially over multiple cycles.}
        \label{fig:adder_serial}
    \end{minipage}%
\end{figure*}
\begin{table}
    \caption{Example of circular shift register state over time for $n_\text{phase bits} = 2$.}
    \label{tab:shift_register}
    \begin{center}
        \begin{tabular}{ccccc}
            \toprule
            \textbf{Time [clock cycles]} & \multicolumn{4}{c}{\textbf{Register Index}} \\
             & \textbf{0} & \textbf{1} & \textbf{2} & \textbf{3} \\ \midrule
            0 & 1 & 1 & 0 & 0 \\ 
            1 & 1 & 0 & 0 & 1 \\
            2 & 0 & 0 & 1 & 1 \\
            3 & 0 & 1 & 1 & 0 \\ 
            4 & 1 & 1 & 0 & 0 \\ \bottomrule
        \end{tabular}
    \end{center}
\end{table}

Figure \ref{fig:adder_parallel} shows a simplified view of the parallel, or combinatorial, implementation of the arithmetic circuits used in the recurrent architecture.
On the left (in green) each of the oscillators are shown.
The middle section (in red) indicates the multiplication by the weights.
However, no actual multiplication is computed since the oscillator amplitude can only have two values, namely 1 and 0, which represent positive and negative amplitude respectively. The weight value or the negative weight value is used for the summation, depending on the amplitude of the respective oscillator.
On the right (in yellow) the summation structure is shown.
It can be seen that for $N$ input values $N - 1$ adders are required.
For a larger number of oscillators the summation structure grows to have a large logic depth, and in turn a long critical path.
In other words, as the number of oscillators goes up, so does the number of adders and thus the length of the logic chain, which means there is a larger delay from the input to the output of the structure.
Theoretically, a smaller, but faster circuit could be created that exploits this large delay to sequentially compute the same sum of all inputs.
This is the inspiration for the hybrid architecture that is discussed in the next section.
Combined with the fact that this arithmetic circuit is replicated for each oscillator, it is easily shown that the number of adders scales proportional to $N^2$.
Hence, the aim is to optimize the implementation of the arithmetic circuit for less hardware resource usage, which is discussed in the next section.

\section{Hybrid ONN Architecture} \label{sec:hybridarchitecture}
The main idea to reduce the hardware resource usage is to utilize a single adder as an accumulator and to serially compute the weighted sum required to update the phase per oscillator.
This is in contrast to the fully parallel nature of the arithmetic circuitry in the recurrent architecture.
However, each oscillator still computes in parallel.
Therefore, this architecture is a hybrid between the recurrent architecture and a fully serial computation, like what would happen in a standard von Neumann architecture.
Since each oscillator will reuse a single adder to compute for each of its connections, instead of one adder per connection, the total number of adders is now proportional to $N$ instead of $N^2$, where $N$ is the number of oscillators.
This is the key difference in architectures that allows for linear resource scaling.

There are a few considerations to maintain equivalent functionality to the recurrent architecture.
Firstly, the final value of the summation should be computed at least as fast as in the recurrent architecture.
In other words, the weighted sum should be computed before the start of the next phase update.
This can be achieved by creating a faster clock domain for the arithmetic logic.
The phase is updated every clock cycle, therefore to serially compute the weighted sum for $N$ oscillators the fast clock domain must be at least $N$ times faster than the existing clock, since at each edge of the fast clock one of the coupling values is processed.
Additionally, the computation must be synchronized to the phase update the final value of the computation must be available before it is needed to update the phase.
The slower clock is used to achieve this.
The rising edge of the slower clock is used to trigger the start of the serial computation for the next rising edge of the slower clock, where the phase will be updated.
That means, when the transition from low a value to high a value on the slow clock is detected, the serial computation will start.
\begin{figure*}
    \centering
    \includegraphics[width=\linewidth]{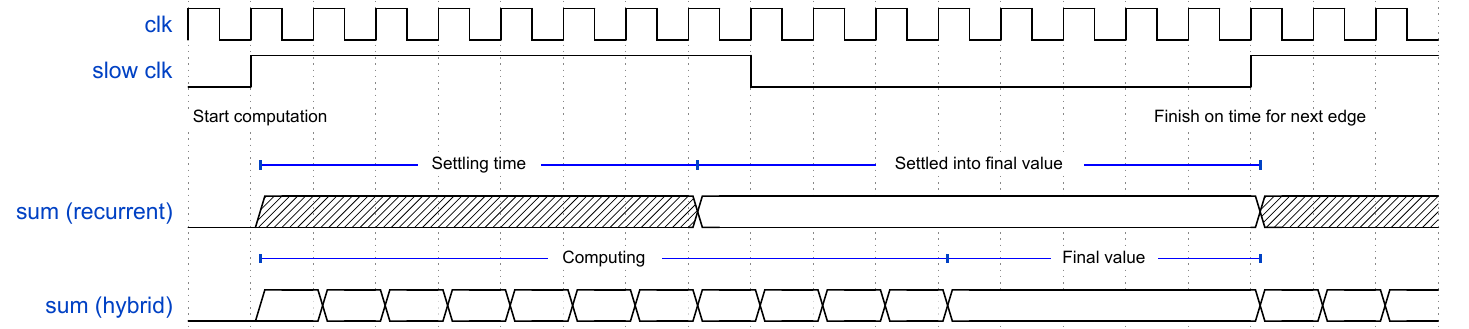}
    \caption{Timing diagram for the computation of a weighted sum for the phase update of one oscillator. The timing for both architectures is shown.}
    \label{fig:timing_diagram}
\end{figure*}
See Figure \ref{fig:timing_diagram} for a timing diagram  of this process.
Figure \ref{fig:adder_serial} shows the serial implementation of the arithmetic circuit.
Again, on the left (in green) the oscillators are pictured.
The output signals of the oscillators will be time multiplexed to the arithmetic circuitry using a multiplexer shown in the middle (in purple).
Since the weights will now be used one-by-one instead of all at the same time, it is possible to store the weights in addressable memory, shown in the right (in red) section above the multiplier block.
During synthesis these memories can be inferred to use the dedicated Block RAM (BRAM) hardware available on the FPGA.
A simple counter is used to select the required memory address and to control the multiplexer.
When the counter has counted up to the number of oscillators, the output of the adder is stored to hold the final sum value while the other components can be reset for the next cycle.
The computation of the sum is now implemented using a single adder that uses feedback from its output to continuously accumulate values.
The second input of the adder is output of the multiplication of the currently selected weight and oscillator output value.
In short the computation will proceed as follows:
At the rising edge of the slower clock the accumulated sum value will be reset to 0.
One-by-one the oscillator amplitudes will be read and multiplied by their corresponding coupling weight.
These values will then be accumulated until the counter reaches the end of computation.
At the end of computation the final value will be stored until the next rising edge of the slower clock, where it will be used to update the phase.

Together, the adder and multiplier constitute a multiply and accumulate operation, which can be inferred to the dedicated digital signal processing (DSP)-slice hardware available on the FPGA.
This frees up more lookup tables and flip-flops to be used for other logic.

\section{Test Setup and Methodology}  \label{sec:methods}
\subsection{Test Platform} \label{sec:methods_platform}
\begin{figure*}
    \centering
    \includegraphics[width=\linewidth]{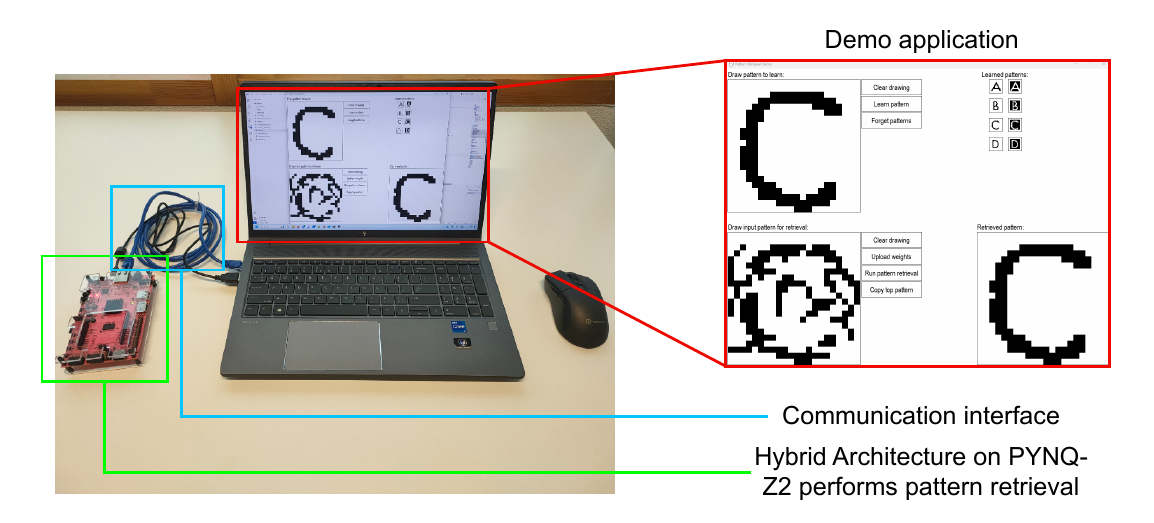}
    \caption{Test setup showing the demonstration user interface and the PYNQ-Z2 performing pattern retrieval using the hybrid architecture.}
    \label{fig:tp_demonstrator}
\end{figure*}

For the following comparisons, a PYNQ-Z2 is used as a test platform.
It features a Zynq-7020 System on Chip with a dual core ARM Cortex-A9 processor and a programmable logic fabric with 85,000 programmable logic cells.
The architectures are written in Verilog using Vivado, which was also used for the hardware synthesis, hardware implementation, and timing analysis.
The pattern retrieval benchmarks are automated using the provided PYNQ Python APIs, which give low-level access to control the hardware through an AXI interface.
Additionally, a demonstration setup was made, which features a Python user interface running on a laptop were a user can draw patterns to memorize and retrieve.
The user interface can communicate with the PYNQ-Z2 over a network connection to transmit the weight matrix and receive the result of the pattern retrieval.
Figure \ref{fig:tp_demonstrator} shows a picture of the demonstration and test setup.

\subsection{Hardware Scaling Analysis Methodology} \label{sec:methods_scaling}
To obtain trends in terms of hardware scaling both architectures were synthesized at different network sizes and the hardware resource usage in terms of lookup tables and flip-flops was noted, since these are the two components determining the amount of logic hardware the architectures need.
For this analysis 5 weight bits and 4 phase bits were used.
In order to get a trend of the trade-off between network size and oscillation frequency, both architectures were also implemented on FPGA to obtain the maximum possible logic frequency, which was then converted to the actual oscillation frequency by the frequency division mechanism discussed earlier.
Additionally, to get a sense of the trade-off between the area utilization and the oscillation frequency we introduce two aggregate measures.
Firstly, the total area used will be defined as the arithmetic mean of the percentages of each FPGA resource used: Flip-Flops, LUTs, DSPs, and BRAMs.
This will give an indication of the percentage of total resources, or FPGA area used.
Secondly, we determine the the percentage of oscillation frequency at each network size of the maximum oscillation frequency achieved.
These two measures will allow us to find a balance point between area and frequency.

For the recurrent architecture the data points for the frequency scaling stop at 48 oscillators due to exceeding the maximum number of lookup tables available on the target FPGA, hence place-and-route could not be completed and by extension timing data could not be extracted.
A standard linear regression was fitted on the base 10 logarithm of the data points to obtain the slope and the $R^2$ value in logarithmic scale.
The slope in the logarithmic scale equals the order of scaling.

\subsection{Pattern Retrieval Methodology} \label{sec:methods_pattern}
The primary goal is to show that the new hybrid architecture performs the same as the recurrent architecture for smaller a real world task up to the maximum implementable size for the test platform.
Additionally, the secondary goal is to show that the new architecture also shows good results for tasks using a larger network size.
If similar performance between the two architectures is shown and high performance can be shown for the larger network sizes, we can be confident that the new architecture does not have significantly different dynamics compared to the original architecture.
Exhaustive benchmarking is not required for this work, as we merely want to verify the functionality of the new architecture.
Therefore, we leave further benchmarking and exploration of applications to future work.
To compare the performance of the two architectures an associative memory task is chosen, namely pattern retrieval from corrupted patterns.
Performance will be measured in retrieval accuracy, meaning the percentage of runs that correctly retrieve the input pattern, and the average time in oscillation cycles to come to the correct solution.
Additionally, the performance of the hybrid architecture is also benchmarked for larger patterns.

There are five datasets, each with different pattern sizes.
The pattern sizes are 3$\times$3, 5$\times$4, 7$\times$6, 10$\times$10, and 22$\times$22.
Note that a 7$\times$6 pattern is the largest size that could be implemented for the recurrent architecture on the test platform, while 22$\times$22 is the largest pattern that could be implemented for the hybrid architecture on the test platform.
Each dataset contains five patterns, representing letters of the alphabet, except the 3$\times$3 dataset, which contains two patterns.
The latter two pattern sizes exceed the maximum number of oscillators available on the recurrent architecture.
To obtain the required coupling weights, each dataset was trained using the Diederich-Opper I learning rule \cite{diederich_learning_1987}.
The resulting weight matrix was quantized to 5 bits signed and programmed into each architecture.
For each dataset, each pattern was corrupted 1000 different times at three different corruption percentages: 10\%, 25\%, and 50\%.
To corrupt a pattern a given percentage of pixels in the pattern was randomly selected and its color was flipped, either from white to black or from black to white.
As an example, corrupting a 10$\times$10 pattern by 10\% means flipping the color on 10 pixels.
Each corrupted pattern was then set as the initial condition for the network after quantizing the phase values to 4 bits.
The final phases were measured and compared to the original target pattern.

\begin{figure}
  \includegraphics[width=\linewidth]{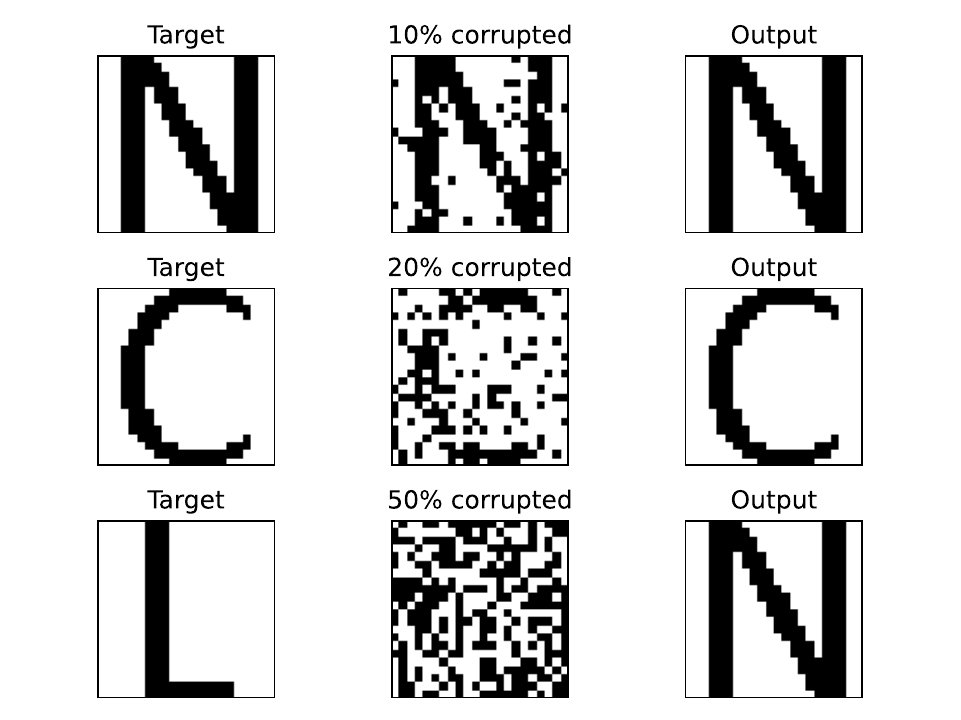}
  \caption{Example of pattern retrieval. In the left column the target images are shown. The center column shows the target images with a given percentage of pixels corrupted. Finally the right column shows the output pattern. It can be seen in the bottom row that the wrong pattern can be retrieved if the input pattern contains too many corrupt pixels.}
  \label{fig:denoising}
\end{figure}

A visual example of the process of retrieving a 22$\times$22 pattern is shown Figure \ref{fig:denoising}.

\section{Scalability Comparison of Architectures} \label{sec:results}
\subsection{Same Numerical Precision} \label{sec:results_same_precision}
\begin{table}
    \caption{Resource usage of each design on a Zynq 7020 FPGA for the maximum feasible number of oscillators using 5 weight bits and 4 phase bits}
    \label{tab:result_resource}
    \begin{center}
        \begin{tabular}{cccc}
            \toprule
            \textbf{Design} & \textbf{Resource} & \textbf{Usage [-]} & \textbf{Usage [\%]}\\ \midrule
            \multirow{4}{*}{Hybrid} & LUT & 41547 & 78.10 \\
             & FF & 44748 & 42.06 \\
             & DSP Slices & 220 & 100 \\
             & Block RAM & 140 & 100 \\ \hline
            \multirow{4}{*}{Recurrent} & LUT & 49441 & 92.9 \\
             & FF & 13906 & 13.1 \\
             & DSP Slices & 0 & 0 \\
             & Block RAM & 0 & 0 \\ 
             \bottomrule
        \end{tabular}
    \end{center}
\end{table}
The architectures will be compared at the same level of numerical precision.
The number of bits were chosen the same as in \cite{abernot_oscillatory_2023}, as it was determined to be sufficient for pattern retrieval using the recurrent architecture derived from that work.
This gives 5 bits for the weight values, including a sign bit, and 4 bits to represent the phase.
At this level of numerical precision the recurrent implementation allowed for a maximum of 48 oscillators, while the serialized implementation achieved 506 oscillators, an increase of factor 10.5.
Table \ref{tab:result_resource} shows the total resource usage for each of the two designs using 5 weight bits and 4 phase bits when synthesized for the maximum number of oscillators possible with the resources available on a Zynq 7020 FPGA.
As can be seen, the limiting factor for the recurrent implementation is the number of lookup tables (LUTs).
The hybrid implementation is limited by the number of DSP slices and Block RAMs on the FPGA.
100\% utilization is not reported due to place-and-route overhead.

\begin{table}
    \caption{Performance of each design on a Zynq 7020 FPGA for the maximum feasible number of oscillators using 5 weight bits and 4 phase bits}
    \label{tab:result_performance}
    \begin{center}
        \begin{tabular}{ccc}
            \toprule
            \textbf{Design} & \textbf{Statistic} & \textbf{Value}\\ \midrule
            \multirow{3}{*}{Hybrid} & Max logic frequency   & 50 MHz    \\
                                    & Oscillation frequency & 6.1 KHz   \\
                                    & Max \#oscillators     & 506       \\
            \hline
            \multirow{3}{*}{Recurrent}  & Max logic frequency   & 40 MHz    \\
                                        & Oscillation frequency & 625 KHz   \\
                                        & Max \#oscillators     & 48        \\
            \bottomrule
        \end{tabular}
    \end{center}
\end{table}

The maximum frequencies and maximum number of oscillators for each architecture are shown in Table \ref{tab:result_performance}.
In this table the logic frequency is the frequency the logic runs at, and oscillation frequency is the frequency of the oscillators after taking into account the frequency division and the length of the shift registers that make up the oscillators.
The recurrent implementation achieved a lower maximum logic frequency compared to the hybrid implementation, but a higher oscillation frequency since the phase update does not need to operate on a slower clock domain.
To allow for the serialization of the computation in the hybrid architecture, this requires additional clock cycles to compute the new weighted sum.
Consequently, the oscillator frequency is slowed down to a lower frequency to allow this new weighted sum update.
Therefore, a trade-off is created between the number of oscillators and the oscillation frequency of the oscillators.
A higher number of oscillators in the hybrid design requires more frequency division to slow down the phase update.
In the next section the scaling of hardware resources and frequency depending on network size will be discussed.

\subsection{Resource Scalability} \label{sec:results_scalability}
\begin{figure}
    \includegraphics[width=\linewidth]{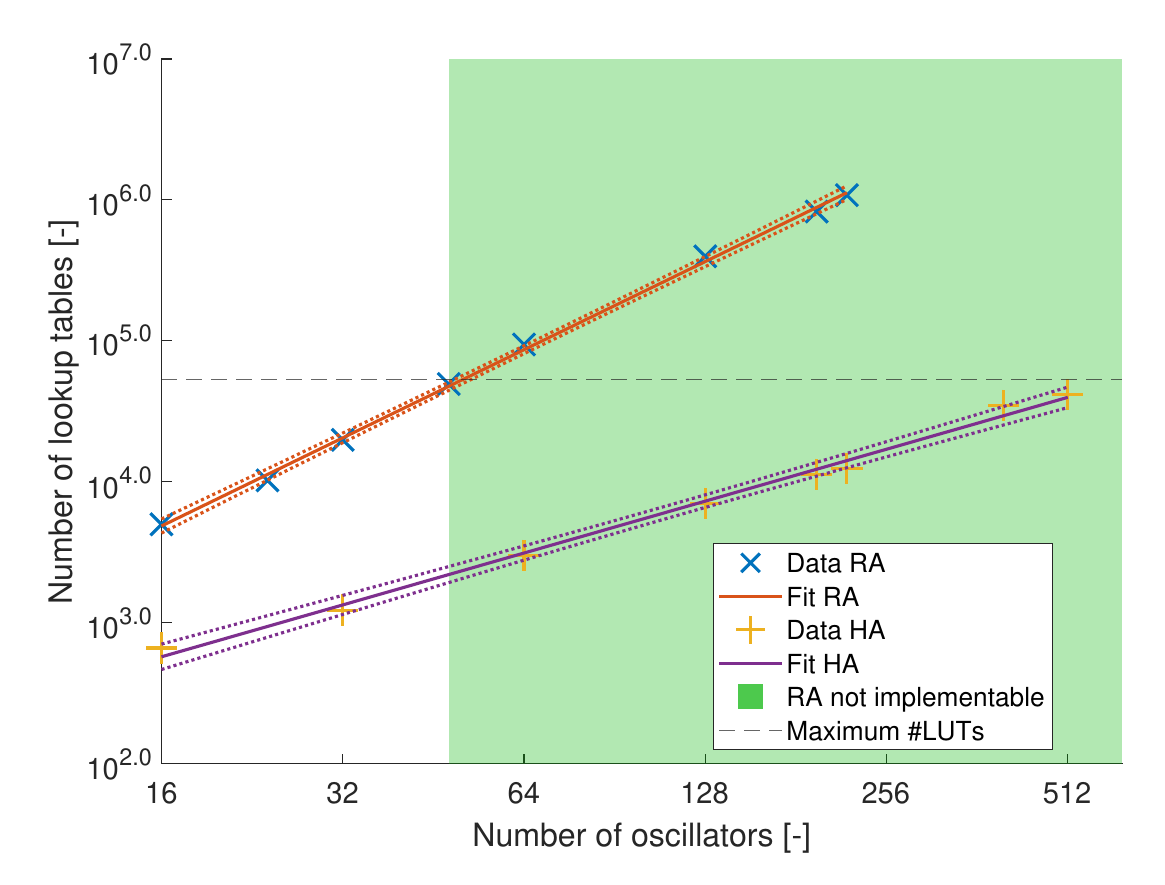}
    \caption{Lookup table (LUT) usage at different network sizes for both architectures at 5 weight bits and 4 phase bits.\\$R^2 = 0.9988$ and slope = $2.0770$ for the recurrent architecture (RA).\\$R^2 = 0.9946$ and slope = $1.2231$ for the hybrid architecture (HA).}
    \label{fig:compare_luts}
\end{figure}

\begin{figure}
    \includegraphics[width=\linewidth]{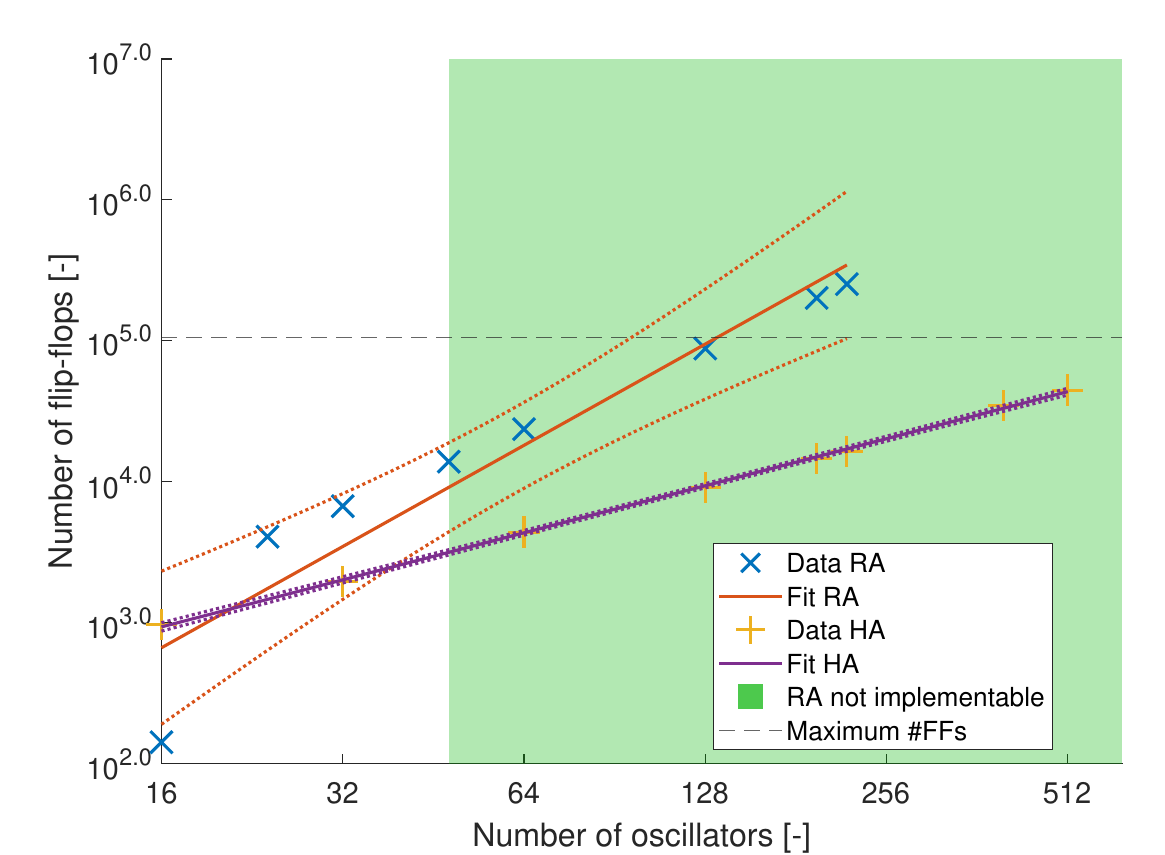}
    \caption{Flip-flop (FF) usage at different network sizes for both architectures at 5 weight bits and 4 phase bits.\\$R^2 = 0.9060$ and slope = $2.3859$ for the recurrent architecture (RA).\\$R^2 = 0.9993$ and slope = $1.1092$ for the hybrid architecture (HA).}
    \label{fig:compare_ffs}
\end{figure}

\begin{figure}
    \includegraphics[width=\linewidth]{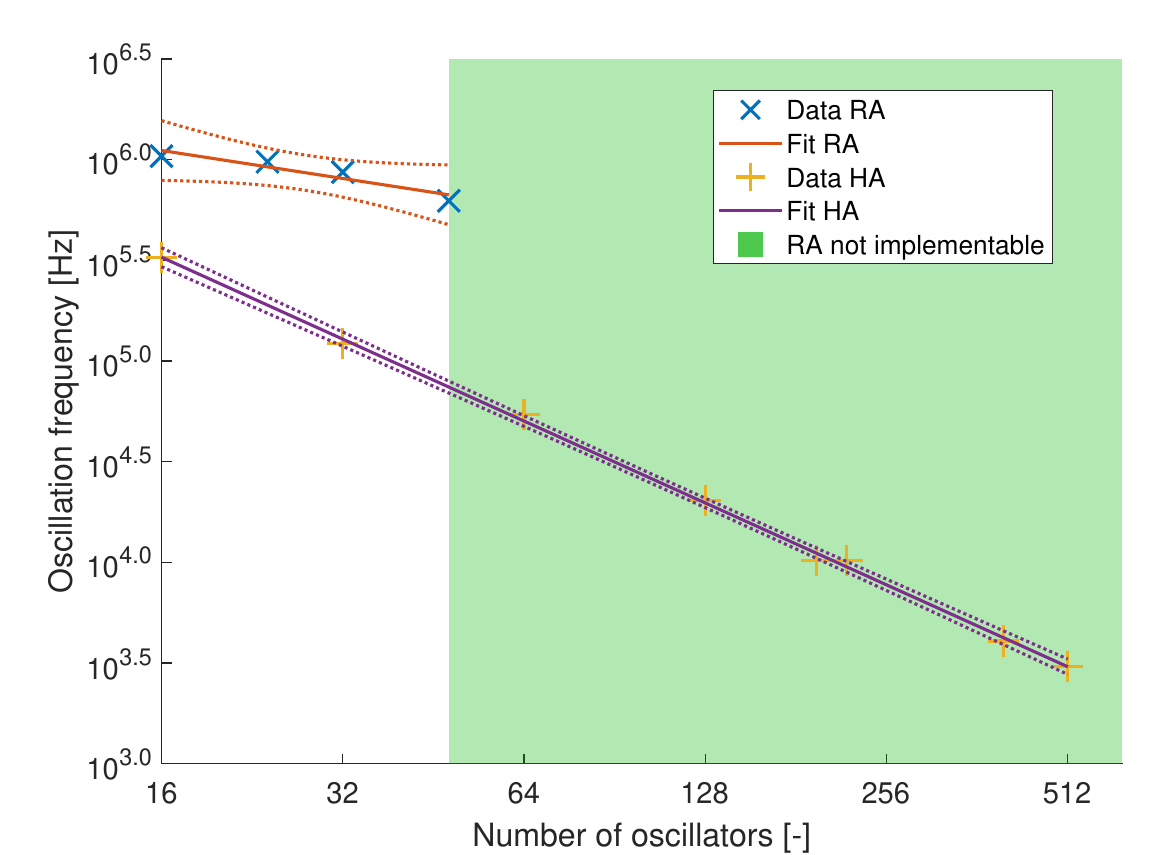}
    \caption{Oscillation frequency at different network sizes for both architectures at 5 weight bits and 4 phase bits.\\$R^2 = 0.8867$ and slope = $-0.4614$ for the recurrent architecture (RA).\\$R^2 = 0.9988$ and slope = $-1.3515$ for the hybrid architecture (HA).}
    \label{fig:compare_freq}
\end{figure}

\begin{figure}
    \includegraphics[width=\linewidth]{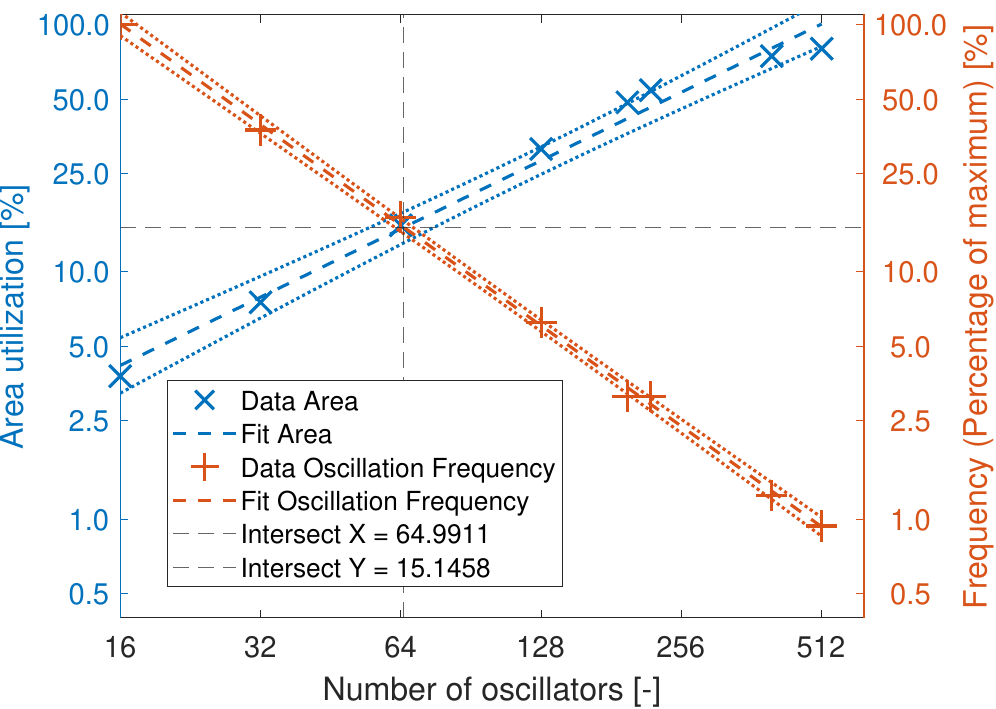}
    \caption{Area utilization and percentage of maximum frequency achieved at different networks sizes for the hybrid architecture.\\
    $R^2 = 0.9852$ for the area utilization.\\
    $R^2 = 0.9988$ for the frequency percentage.\\
    Maximum frequency (100\%) $\approx 325$KHz }
    \label{fig:area_freq_tradeoff}
\end{figure}

Figures \ref{fig:compare_luts}-\ref{fig:compare_freq} show the results of the data analysis for the number of lookup tables versus the number of oscillators, the number of flip-flops versus the number of oscillators, and the oscillation frequency versus the number of oscillators respectively.
The colored area in each figure shows the region where the recurrent architecture could not be implemented.
The dotted lines show the 95\% confidence intervals.
Figure \ref{fig:compare_luts} shows with a high $R^2$ value that the number of lookup tables scales slightly above quadratic for the recurrent architecture, and slightly above linear for the hybrid architecture.
The scaling orders are 2.08 and 1.22 respectively.
Figure \ref{fig:compare_ffs} shows similar results for the number of flip-flops, although the $R^2$ for the recurrent architecture is lower, it is still above 0.9.
Again the recurrent architecture scales above quadratic, having scaling order 2.39.
It can be noted that the data point for the recurrent architecture at 16 oscillators appears to be an outlier and that the true slope might be less steep.
The hybrid architecture scales with order 1.11, again just above linear.
Finally, Figure \ref{fig:compare_freq} shows that the oscillation frequency scales on the order of -0.46, roughly the inverse square root, for the recurrent architecture.
The oscillation frequency of the hybrid architecture scales with order -1.35, which is slightly faster than inversely linear.
Due to the lower number of data points the $R^2$ value for the recurrent architecture is lower, but still relatively high at 0.89.
In Figure \ref{fig:area_freq_tradeoff} the percentage of the total area used is shown alongside the percentage of the maximum oscillation frequency achieved.
We look for a balancing point, which occurs at the intersection of these two lines.
It can be seen in Figure \ref{fig:area_freq_tradeoff} that the intersect is approximately at $N_\text{oscillators} = 65$ and 15\% area utilization.
Figure \ref{fig:area_freq_tradeoff} can also be used to see what network size is feasible at a certain area utilization, or the opposite.
Some applications might demand a certain maximum area usage, while others require a certain frequency.
For example, if an application requires an oscillation frequency of 160KHz, which is around 50\% of the maximum frequency, we can see that this frequency is achieved at approximately 27 oscillators.
At this oscillator count we see that the area usage is roughly 7\%.

\subsection{Pattern Retrieval Results} \label{sec:results_pattern}
The results of the benchmarks are shown in Table \ref{tab:result_denoising_accuracy} and Table \ref{tab:result_denoising_settletime}.
Table \ref{tab:result_denoising_accuracy} shows the retrieval accuracy.
It can be seen that the retrieval accuracy of each architecture are very close or identical for each pattern size and noise level.
The results show that in terms of retrieval accuracy the hybrid architecture performs the same as the recurrent architecture.
This means that the oscillator dynamics of the hybrid architecture are the same as the recurrent architecture.
For larger pattern sizes the hybrid architecture achieves a high performance of 100\% or close to 100\% retrieval accuracy for 10\% and 25\% of pixels corrupted.
It can be concluded from this that the oscillator dynamics do not break down for larger network sizes.
One exception is the retrieval accuracy of the hybrid architecture for 3$\times$3 patterns with 50\% of pixels corrupted.
The accuracy in this case is much higher.
The current hypothesis is that the additional synchronization required in the hybrid architecture slightly changes the system dynamics, which only becomes apparent in the case of a small network with a high noise level.
Note that there exists some run-to-run variance within the results.
This is probably because the signal that enables computation is not synchronized with the oscillators, so the timing of the enable signal related to the oscillation edges has a minor effect on the solution.
\begin{table}
    \caption{Pattern retrieval accuracy for both architectures using 5 weight bits and 4 phase bits.}
    \label{tab:result_denoising_accuracy}
    \begin{center}
        \begin{tabular}{cccc}
            \toprule
            \makecell{\textbf{Pattern}\\\textbf{size}\\\textbf{[pixels]}} & \makecell{\textbf{Corrupted}\\\textbf{pixels}\\\textbf{[\%]}} & \makecell{\textbf{Correctly}\\\textbf{retrieved RA}\\\textbf{[\%]}} & \makecell{\textbf{Correctly}\\\textbf{retrieved HA}\\\textbf{[\%]}}    \\ \midrule
            \multirow{3}{*}{3$\times$3}     & 10 & 100                  & 100   \\
                                            & 25 & 90.8                 & 90.8  \\ 
                                            & 50 & 0                    & 25.8  \\ \hline
            \multirow{3}{*}{5$\times$4}     & 10 & 91.4                 & 91.8  \\ 
                                            & 25 & 50.4                 & 56.0  \\ 
                                            & 50 & 0.3                  & 0.5   \\ \hline
            \multirow{3}{*}{7$\times$6}     & 10 & 99.7                 & 100   \\ 
                                            & 25 & 81.8                 & 89.2  \\ 
                                            & 50 & 0.3                  & 1.0   \\ \hline
            \multirow{3}{*}{10$\times$10}   & 10 & Patterns too large   & 100   \\ 
                                            & 25 & to implement         & 95.4  \\ 
                                            & 50 & on FPGA              & 0.8   \\ \hline
            \multirow{3}{*}{22$\times$22}   & 10 & Patterns too large   & 100   \\ 
                                            & 25 & to implement         & 100   \\ 
                                            & 50 & on FPGA              & 0     \\ \bottomrule
        \end{tabular}
    \end{center}
\end{table}

Table \ref{tab:result_denoising_settletime} shows the arithmetic mean of the runtime in terms of oscillation cycles excluding time-outs.
Similar to the retrieval accuracy, it can be concluded that the time to settle of both architectures is within margin of error.
The settling times for the two larger pattern sizes on the hybrid architecture do not explode compared to smaller sizes.
It is noted, that although the time to settle in terms of oscillation cycles is the same, the time to settle in terms of real time is higher for the hybrid architecture, due to the lower oscillation frequency.
\begin{table}
    \caption{Arithmetic mean pattern retrieval time to settle for both architectures using 5 weight bits and 4 phase bits, excluding time-outs}
    \label{tab:result_denoising_settletime}
    \begin{center}
        \begin{tabular}{cccc}
            \toprule
            \makecell{\textbf{Pattern}\\\textbf{size}} & \makecell{\textbf{Corrupted}\\\textbf{pixels}\\\textbf{[\%]}} & \makecell{\textbf{Mean time}\\\textbf{to settle RA}\\\textbf{[cycles]}} & \makecell{\textbf{Mean time}\\\textbf{to settle HA}\\\textbf{[cycles]}} \\ \midrule
            \multirow{3}{*}{3$\times$3}     & 10 & 10.1                 & 10.1      \\
                                            & 25 & 10.2                 & 10.1      \\ 
                                            & 50 & 11.1                 & 11.7      \\ \hline
            \multirow{3}{*}{5$\times$4}     & 10 & 19.6                 & 19.8      \\ 
                                            & 25 & 22.9                 & 23.8      \\ 
                                            & 50 & 25.6                 & 26.5      \\ \hline
            \multirow{3}{*}{7$\times$6}     & 10 & 26.1                 & 25.8      \\ 
                                            & 25 & 30.6                 & 28.6      \\ 
                                            & 50 & 36.3                 & 32.6      \\ \hline
            \multirow{3}{*}{10$\times$10}   & 10 & Patterns too large   & 25.5      \\ 
                                            & 25 & to implement         & 27.0      \\ 
                                            & 50 & on FPGA              & 32.6      \\ \hline
            \multirow{3}{*}{22$\times$22}   & 10 & Patterns too large   & 25.5      \\ 
                                            & 25 & to implement         & 25.5      \\ 
                                            & 50 & on FPGA              & 33.3      \\ \bottomrule
        \end{tabular}
    \end{center}
\end{table}

To summarize, the hybrid and recurrent architecture perform close to identical in terms of retrieval accuracy and settling time and the hybrid architecture shows good performance for larger patterns.
But the proposed architecture achieves these results for less hardware resources used and allows for larger implementation by achieving near-linear hardware scaling.

\section{Discussion} \label{sec:discussion}
This paper has shown a new hybrid architecture based on a previous recurrent digital ONN \cite{abernot_digital_2021, abernot_oscillatory_2023, abernot_training_2023, luhulima_digital_2023}.
It was analyzed in terms of hardware resource scaling, where it showed a near-linear scaling in contrast to the previous quadratic scaling trend.
This near-linear scaling allows the hardware to be implemented with much larger ONN sizes.
However, this near-linear scaling comes at the cost of oscillation frequency.
If a high oscillation frequency is required in a certain application, a user might choose a lower number of oscillators with a higher oscillation frequency.

For the specific test platform used in this paper, the ONN size could be increased by 10.5$\times$ on a single FPGA.
In future work, even larger network sizes could be achieved by clustering multiple FPGAs, however synchronizing multiple ONNs across multiple devices will pose a challenge.
Larger network sizes can be benchmarked using datasets that require more nodes to be embedded on ONNs, especially combinatorial optimization problems.
Additionally, the new architecture was verified against the existing architecture by means of an associative memory task.
Here it was shown that the hybrid architecture achieved similar retrieval accuracies as the recurrent architecture across a dataset.
The time to settle in terms of oscillation cycles was also similar between the two architectures.
Now that this level of scalability has been achieved, the next step is to explore real-world applications.
Future work can compare the architecture introduced in this work to other computing paradigms using other benchmarks and applications.

\section{Conclusion} \label{sec:conclusion}
A hybrid architecture based on an existing recurrent digital ONN architecture has been presented, which allows for a 10.5 times increase in the number of oscillators to be implemented on FPGA at the same level of numerical precision.
In this architecture part of the phase update computation was serialized to save on arithmetic logic.
However, serialization of the computation created a trade off between network size and oscillation frequency.
The oscillation frequency reduces by an order of 1.35 for the number of oscillators in the network.
An analysis of hardware resource scaling was performed that showed that the recurrent architecture has hardware resource usage that scales roughly quadratically.
The same analysis showed that the hybrid architecture has hardware resource usage that scales nearly linear, on the order of about 1.2.
Furthermore, a pattern retrieval benchmark against the recurrent architecture was performed to verify the performance of the new architecture.
Similar results in terms of retrieval accuracy and retrieval time has been found for both architectures, indicating the presence of similar dynamics in both architectures.
Additionally, retrieval accuracies of 100\% have been shown for larger pattern retrieval task that are made possible by the more efficient hardware resource usage of the hybrid architecture.
To the best of our knowledge, this work presents the largest fully connected digital ONN architecture in terms of number of oscillators implemented thus far and an architecture that achieves nearly linear hardware scaling.


\bibliographystyle{ACM-Reference-Format}
\bibliography{library_b_haverkort}



\end{document}